\documentclass[submission,copyright,creativecommons]{eptcs}
\providecommand{\event}{          
    Horn Clause Verification and Synthesis (HCVS 2019)
}
\usepackage{breakurl}             
\usepackage[strings]{underscore}  
\usepackage{color}

\newcommand{\tdagger}{\dagger\hspace*{-1mm}\dagger}

\title{Proving Properties of Sorting Programs:\\
A Case Study in Horn Clause Verification}

\author{
Emanuele De Angelis
\institute{DEC, University ``G.~d'Annunzio" of Chieti-Pescara\\[2pt]
    Viale Pindaro 42, 65127 Pescara, Italy
}
\email{emanuele.deangelis@unich.it}
\and
Fabio Fioravanti
\institute{DEC, University ``G.~d'Annunzio" of Chieti-Pescara\\[2pt]
    Viale Pindaro 42, 65127 Pescara, Italy
}
\email{fabio.fioravanti@unich.it}
\and
Alberto Pettorossi
\institute{
    DICII, University of Roma Tor Vergata\\[2pt]
    Via del Politecnico 1, 00133 Roma, Italy
}
\email{pettorossi@info.uniroma2.it}
\and
Maurizio Proietti
\institute{
    CNR-IASI\\[2pt]
    Via dei Taurini 19, 00185 Roma, Italy
}
\email{maurizio.proietti@iasi.cnr.it}
}

\def\titlerunning{Proving Properties of Sorting Programs: 
	A Case Study in Horn Clause Verification}

\def\authorrunning{E.~De Angelis et al.}

\begin{document}
\maketitle

\begin{abstract}
\begin{abstract}
The proof of a program property 
can be reduced to the proof of
satisfiability of a set of constrained Horn clauses (CHCs)
which can be automatically generated from the program and the property.
In this paper we have conducted a case study in Horn clause verification by 
considering several sorting
programs with the aim of exploring the effectiveness of a 
transformation technique 
which allows us to eliminate 
inductive data structures such as lists or trees. 
If this technique is successful, we derive a set of CHCs 
with constraints over the integers and booleans only,
and  the satisfiability check can often be performed in an effective way
by using state-of-the-art CHC solvers, 
such as Eldarica or~Z3. 
In this case study we have also illustrated the
usefulness of a companion technique based on the 
introduction of the so-called {\it{difference}} \mbox{\it{predicates}},
whose definitions correspond to lemmata required 
during the verification.  
We have considered functional programs which implement 
the following kinds of sorting algorithms acting on lists of integers: 
(i)~linearly
 recursive sorting algorithms, such as insertion sort and selection sort, 
and (ii)~non-linearly recursive sorting 
algorithms, such as quicksort and mergesort, and
we have considered the following properties: 
(i)~the partial correctness properties, that is, the  
orderedness of the output lists, and the equality of 
the input and output lists when viewed as multisets, and  
(ii)~some arithmetic properties, such as 
the equality of the sum of the elements before and after sorting.
\end{abstract}

\end{abstract}

\section{Introduction}
\label{intro}

Recent work has shown that {\em{constrained Horn clauses}} (CHCs)
provide a suitable basis for program verification~\cite{Bj&15},
and indeed many program verification problems can be reduced 
in a very natural way to CHC satisfiability problems. 
This fact has motivated the development of a number of tools, 
called {\em CHC solvers}, 
which can perform satisfiability proofs. Most of them 
work well on CHCs with simple constraint theories, such as the 
theory of Linear Integer Arithmetic ($\mathrm{LIA}$) and the theory 
of Booleans~($\mathrm{Bool}$). Among these tools we mention:
Eldarica~\cite{HoR18}, \mbox{FreqHorn~\cite{Fe&18},}
HSF~\cite{Gr&12}, RAHFT~\cite{Ka&16}, Spacer~\cite{Ko&14},
VeriMAP~\cite{De&14b}, and Z3~\cite{DeB08}. 

Unfortunately, when the properties to be verified refer to programs 
that act on inductively 
defined data structures, such as lists or trees, then 
the satisfiability proofs via CHC solvers becomes much harder, 
or even impossible, because those solvers do not natively incorporate 
induction principles relative to those data structures.

In order to overcome this difficulty, the following two approaches 
have recently been suggested.
(i)~The first one consists in the
incorporation into CHC solvers of suitable
induction principles~\cite{ReK15,Un&17}, and 
(ii)~the second one consists in transforming the given set of CHCs 
into a new set where inductively defined data structures are removed, and whose satisfiability
implies the satisfiability of the original clauses~\cite{De&18a,MoF17}.

In this paper we will follow this second approach and, 
in particular, we will consider the 
{\em Elimination Algorithm} presented in a previous work of ours~\cite{De&18a},
which implements a transformation strategy for removing inductively defined data
structures and it is 
based on the familiar fold/unfold rules~\cite{EtG96,TaS84}.
Thus, if the clauses derived by the Elimination Algorithm have all 
their constraints 
in the {\rm{LIA}} or {\rm{Bool}} theory, then
there is no need to modify the CHC solvers for performing 
the required satisfiability proofs.

As usual in the transformation-based approach, 
the success of the Elimination Algorithm depends on the introduction of
some suitable auxiliary predicates. 

The novel contribution of this paper is the study of various 
verification problems of sorting programs which we have verified by using
a technique for introducing auxiliary
predicates, called {\it the introduction of difference predicates}~\cite{De&19a}. 
The name of the difference predicates comes from the fact that they
express the relation between the values computed by two different functions.
The definitions of those predicates also
correspond to 
the lemmata required for the proofs of
the properties of interest by structural induction.

By extending the Elimination Algorithm
with the introduction of difference predicates, we can remove inductively 
defined data structures
in cases where the plain Elimination Algorithm would not terminate. 
By doing so we extend the effectiveness of state-of-the-art CHC 
solvers when proving program properties.

In Sections~\ref{LinearRecursionVerification} and~\ref{NonLinearRecursionVerification},
we will consider various verification problems of sorting algorithms
written as functional programs acting on lists of integers.
In particular in Section~\ref{LinearRecursionVerification} we deal with linearly recursive
sorting algorithms, while in Section~\ref{NonLinearRecursionVerification} we deal with 
non-linearly recursive ones. The verification
is done into two phases. In Phase~(i) we derive 
by transformation, introducing suitable difference predicates,
a set of CHCs on LIA and Bool constraints only (that is, constraints on lists
will no longer be present in the derived clauses), and then in
Phase~(ii) we prove the satisfiability of the derived CHCs 
by using the solver Eldarica acting on LIA and Bool constraints.
We have also proved that Eldarica is not able (within a given time-out)
to check satisfiability of 
the clauses which are obtained by the direct translation into CHCs
of the given functional programs and properties, before the 
transformation performed at Phase~(i).
In Section~\ref{ConcludingRemarks} we present the related work and we make
some concluding remarks. In the Appendix we provide the technical details
of the verifications we have performed by using our verification and
transformation system VeriMAP~\cite{De&14b} and the CHC 
solver Eldarica~\cite{HoR18}.

\section{Verification of Linearly Recursive Sorting Algorithms}
\label{LinearRecursionVerification}
\newcommand{\bbar}{${\mathtt |\hspace{-.0mm}|}$}

In this section, we consider two  functional programs
implementing simple linearly recursive sorting algorithms: 
{\em InsertionSort} and {\em SelectionSort}.
Then, we verify various properties of those
programs by applying our transformation-based method.
The transformations presented here and in the next section are done by 
an interactive system\,\footnote{The MAP Transformation System \url{http://www.iasi.cnr.it/~proietti/system.html}}.
Their mechanization is briefly discussed in Section~\ref{ConcludingRemarks}.

Let us consider the following  {\it InsertionSort} program 
written in OCaml syntax~\cite{Le&17}:

\vspace{-1mm}

\begin{samepage}
\begin{verbatim}
type list = Nil | Cons of int * list;;
let rec ins x l =
  match l with
  | Nil -> Cons(x,Nil)
  | Cons(y,ys) -> if x<=y then Cons(x,Cons(y,ys)) else Cons(y,ins x ys);;
let rec insertionSort l =
  match l with
  | Nil -> Nil
  | Cons(x,xs) -> ins x (insertionSort xs);;
let rec count x l =
  match l with
  | Nil -> 0
  | Cons(y,ys) -> if x=y then 1 + count x ys else count x ys;;
\end{verbatim}
\end{samepage}

\vspace{-1mm}

\noindent
where: (i)~given an integer {\tt x} and a list {\tt l} of integers, 
ordered in the ascending order, {\tt ins x l} 
returns the ordered list made out of {\tt x} and the 
integers in~{\tt l},
(ii)~{\tt insertionSort l} returns an ordered permutation 
of the integers in the list~{\tt l}, and 
(iii)~{\tt count x l\,} returns the number of occurrences of the integer {\tt x} in 
the (not necessarily ordered) list {\tt l}.

Let us suppose that we want to prove the following 
property of 
{\it InsertionSort\,}: 

$\mathtt{\forall\, l,s,x.}$ {\tt (insertionSort l)$\,=\,$s}   
~~~$\rightarrow$~~~ {\tt(count x l)$\,=\,$(count x s)}
\hfill $(\mathit{IS\_Perm})$\hspace{10mm}

\noindent
which states that the list {\tt s} returned by {\tt insertionSort l} 
is a permutation of the given list {\tt l}.

In order to prove this property, we first consider the translation 
of  {\it InsertionSort} 
and  $\mathit{IS\_Perm}$ into the following set of 
constrained Horn clauses (we will not enter into the details of this translation here 
and the reader may refer to~\cite{De&18a,Un&17})\,\footnote{When writing clauses we use 
a Prolog-like syntax, instead of the more verbose SMT-LIB syntax.}:

\vspace{-1mm}
\begin{verbatim}
1. false :- N1=\=N2, insertionSort(L,S), count(X,L,N1), count(X,S,N2).
2. ins(I,[],[I]).
3. ins(I,[X|Xs],[I,X|Xs]) :- I=<X.
4. ins(I,[X|Xs],[X|Ys]) :- I>X, ins(I,Xs,Ys).
5. insertionSort([],[]). 
6. insertionSort([X|Xs],S) :- insertionSort(Xs,S1), ins(X,S1,S).
7. count(X,[],0).
8. count(X,[H|T],N) :- X=H,  N=M+1, count(X,T,M).
9. count(X,[H|T],N) :- X=\=H, count(X,T,N).
\end{verbatim}

\vspace{-1mm}

\noindent 
The translation ensures that (i)~{\tt ins(X,L,L1)}, 
(ii)~{\tt{insertionSort(L,S)}}, and (iii)~{\tt count(X,L,N)} hold iff 
(i)~{\tt ins~X~L = L1}, (ii)~{\tt{insertionSort~L = S}}, and (iii)~{\tt count~X~L = N},
respectively, hold in the program {\it InsertionSort}.
Clause {\tt1} translates property~$\mathit{IS\_Perm}$ as it stands
(using the functional notation)  for:

\smallskip

\noindent
~~~$\mathtt{\forall\, l,s,x,n1,n2.}$ {\tt (insertionSort~l)} $\mathtt{= s 
~\wedge {(count}~x~l)\! =\! n1 ~\wedge~ ({count}~x~s)\! =\! n2 
  ~~\rightarrow~~ {n1\!=\! n2}} $

\smallskip

\noindent
Now, as it is well-known from the literature, we have that clauses~{\tt1}--{\tt9} are satisfiable iff ${\mathit{IS\_Perm}}$ holds. 
Unfortunately, state-of-the-art CHC solvers, such as Eldarica or Z3, 
fail to prove satisfiability of those clauses,
because these solvers do not incorporate any 
induction principle on lists. 

Thus, we proceed by applying a transformation technique 
based on the Elimination Algorithm~\cite{De&18a} whose objective is to
eliminate the inductively defined data structures and, in particular, in our case,
 the lists. 
That algorithm is successful
only if it is combined with a companion technique for
the introduction of the so-called {\it difference predicates} whose application
we will now illustrate. 

At the end of the transformation, if the derived clauses have 
no occurrences of lists,
we can check their satisfiability by using a state-of-the-art CHC solver
on LIA. 
From the proof of satisfiability, we will then get
the proof of the property {\it IS\_Perm}
because during the transformation every rule that has been applied 
preserves satisfiability~\cite{EtG96}.

Let us start off by applying the {Elimination Algorithm} to clause~{\tt1}. 
We introduce a new predicate {\tt new1} defined by the following clause
(here and in what follows the $n$ atoms in the body 
of clause~{\tt m} are identified by numbers {\tt m.1}, $\ldots$, {\tt m.n}, respectively):

\vspace{-1mm}

\begin{verbatim}
10. new1(X,N1,N2) :- insertionSort(L,S), count(X,L,N1), count(X,S,N2).
                          (10.1)             (10.2)          (10.3)
\end{verbatim}
\vspace{-1mm}

\noindent 
The atoms in the body of clause~{\tt10} are the atoms in the body
of clause~{\tt1} and the arguments of {\tt new1} are the non-list variables 
occurring in those atoms (this choice is consistent with our objective of eliminating
the list variables). 
We fold clause~{\tt 1} using clause~{\tt10} and we get the following clause
with no list variables:

\vspace{-1mm}

\begin{verbatim}
11. false :- N1=\=N2, new1(X,N1,N2).
\end{verbatim}

\vspace{-1mm}

\noindent
Now we proceed by eliminating lists from the newly introduced clause~{\tt10}. We unfold 
it and we get:

\vspace{-2mm}

\begin{verbatim}
12. new1(X,0,0).
13. new1(X,N1,N2) :- N1=N+1, insertionSort(Xs,S1), ins(X,S1,S),
                                  (13.1)               (13.2)
                     count(X,Xs,N), count(X,S,N2).
                        (13.3)        (13.4)
14. new1(X,N1,N2) :- X=\=Y, insertionSort(Xs,S1), ins(Y,S1,S), 
                     count(X,Xs,N1), count(X,S,N2).
\end{verbatim}

\vspace{-2mm}

\noindent 
At this point it is impossible to fold clauses~{\tt13} and~{\tt14}
using clause~{\tt10}
because of the mismatch of the variables~{\tt S} and~{\tt S1}. Thus, 
we apply the {difference predicate} technique which works according to the
following Steps~1--6.
Let us consider clause~{\tt13}. Analogous steps can be performed for
clause~{\tt14} and we leave them to the reader.

\smallskip
\noindent
$\bullet$ {\it Step} 1. {\it Embed}. Each of the atoms~{\tt13.1}, {\tt13.3}, 
and {\tt13.4} of clause~{\tt 13} to be folded is a variant of atom~{\tt10.1}, 
{\tt10.2} and {\tt10.3}, respectively, of clause~{\tt 10} to be used for folding. 
We say that clause~{\tt10} {\it is embedded into} clause~{\tt13}.
However, their conjunction is {\it not} a 
variant of the body of clause~{\tt10}, and thus we cannot fold clause~{\tt 13}
using clause~{\tt10}.

\smallskip
\noindent
$\bullet$ {\it Step} 2. {\it Rename}. We rename apart clause {\tt10} to be used for folding, 
so as to have fresh, new variable names that do not occur anywhere else. 
This renaming apart 
will avoid clashes of variables in all the subsequent steps. We get:

\vspace{-1mm}

\begin{small}
\begin{verbatim}
10a. new1(Xa,N1a,N2a) :- insertionSort(La,Sa), count(Xa,La,N1a), count(Xa,Sa,N2a).                            
                              (10a.1)              (10a.2)           (10a.3)
\end{verbatim}
\end{small}

\vspace{-1mm}

\noindent
$\bullet$ {\it Step} 3. {\it Match}. We match the body of clause~{\tt10a}  against 
the body of clause~{\tt13} to be folded. We manage to match the conjunction ({\tt10a.1}, 
{\tt10a.2}) with 
the conjunction ({\tt13.1}, {\tt13.3}) by the renaming substitution 
$\sigma= \{{\mathtt{La/Xs,}}$ ${\mathtt{Sa/S1,}}$ 
${\mathtt{Xa/X,}}$ ${\mathtt{N1a/N}}\}$, 
but we cannot extend this matching
to the remaining atoms {\tt10a.3} and {\tt13.4}
because the substitution $\{{\mathtt{Xa/X,~Sa/S,~N2a/N2}}\}$ 
is inconsistent with $\sigma$. 
By applying the substitution~$\sigma$, we get the following clause~{\tt10m}, which 
is a variant of clause~{\tt10a}:

\smallskip
\noindent
${\mathtt{10m.~~new1(X,N,N2a)~:\mbox{\tt -}~insertionSort(Xs,S1),~count(X,Xs,N),
   ~||~count(X,S1,N2a).}}$ 
   
\noindent
${\mathtt{\hspace{52mm}(10m.1) \hspace{24mm} (10m.2)\hspace{18mm}(10m.3)}}$    

\smallskip
\noindent
This clause~{\tt 10m} is the actual
clause which we will use for folding at Step~6 below.
The marker~\bbar~in the body of clause {\tt10m} has no logical meaning and it
is only used for separating the {\it matching conjunction} 
\mbox{(atoms {\tt10m.1} and 
{\tt10m.2})} to its left and the {\it mismatching conjunction} 
\mbox{(atom {\tt10m.3})}
to its right. In general, also the {\rm mismatching conjunction} may 
consist of more than
one atom, although in our case it is made out of one atom only.
Also for the clause to be folded (clause~{\tt13} in our case)
we define the matching and the mismatching conjunctions:
(i)~the {\it matching conjunction} 
is equal to the one of the clause
we will use for folding (atoms~{\tt 13.1} and~{\tt 13.3} in our case),
while (ii)~the {\it mismatching conjunction} 
is made out of all body atoms of the clause to be folded
that do not belong
to the {\rm matching conjunction} (atoms~{\tt13.2} and~{\tt13.4} in our case).


\smallskip
\noindent
$\bullet$ {\it Step} 4. {\it Introduce a Difference Predicate}. 
Now, in order to fold 
clause~{\tt13} using clause~{\tt 10m} we need to replace 
the mismatching conjunction of clause~{\tt13}
by the mismatching conjunction of clause~{\tt10m}. This replacement 
can be done at the expense of adding to the body of clause~{\tt 13} a new atom
with a so-called {\it difference predicate}. 
This atom addition is required for preserving satisfiability of the derived
clauses. 
In our case the difference predicate we introduce, called {\tt diff1}, is 
defined as follows: 

\smallskip
\noindent
${\mathtt{15.~~diff1(X,N2a,N2)~:\mbox{\tt -}~ins(X,S1,S),~count(X,S,N2),}}$ $\tdagger$
${\mathtt{count(X,S1,N2a).}}$ 
   
\noindent
${\mathtt{\hspace{50mm}(13.2) \hspace{12mm} (13.4)\hspace{22mm}(10m.3)}}$                                                 
\smallskip

\noindent
This definition clause is constructed as we now specify.
The body is made out of two conjunctions separated by the marker~$\tdagger$: 
(i)~the first one to the left of~$\tdagger$ is the mismatching conjunction of 
the clause to be folded (atoms~{\tt13.2} and ~{\tt13.4} in our case),
and 
(ii)~the second conjunction to the right of~$\tdagger$ is the mismatching conjunction of the
clause we will use for folding (atom {\tt 10m.3} in our case).
The arguments~{\tt X}, {\tt N2a}, and {\tt N2} of the predicate 
{\tt diff1} are the non-list variables 
occurring in the body we have just constructed 
(obviously these arguments can be placed in any order).
The marker~$\tdagger$~in clause~{\tt15} separates
the atoms to its left that should be removed from the body of clause~{\tt 13} from 
the atoms to its right that should be added to the body of clause~{\tt 13}
so that folding of clause~{\tt13} can be performed as we desire (see Step~6 below).

The reader may note that the difference predicate {\tt diff1(X,N2a,N2)}
expresses the relation between
the non-list output variable~{\tt N2a} of the atom that is added and 
the non-list output variable~{\tt N2} of the atoms that are 
removed~\footnote{~The input and output variables 
of the atoms {\tt10m.3}, {\tt13.2}, and {\tt13.4} are defined as expected, 
if one considers the associated functional expressions \mbox{\tt (count~X~S1)}, 
\mbox{\tt (ins~X~S1)}, and \mbox{\tt (count~X~S)}, respectively.}.

\smallskip
\noindent
$\bullet$ {\it Step} 5. {\it Replace}. In the clause to be folded 
(in our case clause~{\tt13}) we replace the mismatching 
conjunction 
(in our case atoms~{\tt13.2} and~{\tt13.4}) by: 
(i) the mismatching conjunction  of the clause we
will use for folding (that is, atom {\tt 10m.3)}, and 
(ii)~the head of
the definition of the difference predicate~{\tt diff1}. 
Note that no extra variable renamings are required besides that of 
Step~(2).
We get the following clause:

\vspace{-1mm}

\begin{verbatim}
13r. new1(X,N1,N2) :- N1=N+1, insertionSort(Xs,S1), count(X,Xs,N), 
                                    (13.1)             (13.3)                       
                      count(X,S1,N2a), diff1(X,N2a,N2). 
                           (10m.3)    
\end{verbatim}

\vspace{-1mm}

\noindent
$\bullet$ {\it Step} 6. {\it Fold}. We fold clause~{\tt13r} 
using clause~{\tt10m}
(this folding is possible by construction) and we get:

\vspace{-1mm}

\begin{verbatim}
13f. new1(X,N1,N2) :- N1=N+1, new1(X,N,N2a), diff1(X,N2a,N2).  
\end{verbatim}

\vspace{-1mm}

\noindent
It can be shown that the above Steps 1--6 preserve
satisfiability of clauses, and this is an essential requirement 
in our proof of the property {\it IS\_Perm}. A detailed proof
of this fact is outside the scope of the present paper.

Analogous steps can be performed starting from clause~{\tt14},
and during these steps we introduce a new difference predicate
{\tt diff2} defined by the following clause:

\vspace{-1mm}

\begin{verbatim}
16. diff2(X,Y,N2b,N2) :- X=\=Y, ins(X,S1,S), count(Y,S,N2), count(Y,S1,N2b).
\end{verbatim}
                                       
\vspace{-1mm}

\noindent 
Note that, unlike clause~{\tt{15}} 
defining the predicate {\tt{diff1}}, clause~{\tt{16}} has in its body
also the constraint {\tt{X=$\backslash$=Y}}. We have to add this constraint
because: (i)~it occurs in the
body of the clause to be folded (clause~{\tt 14}, in our case), and 
(ii)~it relates variables ({\tt{X}} and {\tt{Y}}, in our case) occurring 
in the body atoms.

The reader may check that, as from clause~{\tt13} we derived clause~{\tt13f}, from clause~{\tt 14} 
we may derive the following folded clause:

\vspace{-1mm}

\begin{verbatim}
14f. new1(X,N1,N2) :- X=\=Y, new1(Y,N1,N2b), diff2(X,Y,N2b,N2).
\end{verbatim}

\vspace{-1mm}

\noindent
The CHCs we have obtained so far are: {\tt11}, {\tt12}, {\tt13f}, {\tt14f},
together with: (i)~clauses~{\tt15} and~{\tt16} defining 
the two difference 
predicates~{\tt diff1} and~{\tt diff2}, respectively, and (ii)~clauses~{\tt2}--{\tt4} and
{\tt7}--{\tt9}
defining the predicates {\tt ins} and {\tt count}. 
Since clauses~{\tt15} and~{\tt16} 
still have list variables in their bodies, we continue the application of
the Elimination Algorithm~\cite{De&18a} with the objective of eliminating list
variables. During this process, 
which we will not describe here, no new difference predicates
need to be introduced, and we get the following final set of clauses, besides
 clauses~{\tt11}, {\tt12}, {\tt13f}, {\tt14f}:

\vspace{-1mm}

\begin{verbatim}
diff1(X,0,N2)  :- N2=N1+1, new2(X,N1).
diff1(X,N1,N2) :- N2=M2+1, N1=M1+1, new3(X,M2,M1), 
diff1(X,N1,N2) :- X=<Y, N2=N+1, X=\=Y, new4(X,Y,N,N1).
diff2(X,Y,0,0) :- Y=\=X.
diff2(X,Y,M,N) :- X=<Y, Y=\=X, M=K+1, new3(Y,N,K).
diff2(X,Y,M,N) :- X=<Z, Y=\=X, Y=\=Z, N=M, new5(Y,N).   
diff2(X,Y,M,N) :- X>Y, N=H+1, M=K+1, diff2(X,Y,K,H). 
new2(X,0).
new3(X,N1,N) :- N1=N+1, new5(X,N).
new4(X,Y,N,N) :- X=<Y, X=\=Y, new5(X,N).
new5(X,0).
new5(X,N1) :- N1=N+1, new5(X,N).
\end{verbatim}

\vspace{-1mm}

\noindent
If we submit this final set of clauses to the CHC solver Eldarica~\cite{Ho&12}, it 
is able to show its 
satisfiability by constructing the following model (we present it using
 Eldarica's {\em Prolog format}\/):

\vspace{-1mm}

\begin{verbatim}
diff1(A,B,C) :- (((B - C) = -1), (B >= 0)).
diff2(A,B,C,D) :- ((C = D), (C >= 0)).
new1(A,B,C) :- ((B = C), (B >= 0)).
new2(A,B) :- (B = 0).
new3(A,B,C) :- (((C - B) = -1), (B >= 1)).
new4(A,B,C,D) :- ((D = C), ((C >= 0), ((B - A) >= 1))).
new5(A,B) :- (B >= 0).
\end{verbatim}

\vspace{-1mm}

\noindent
The existence of this model shows that property {\it IS\_Perm} holds
 for program {\it InsertionSort}.

As already mentioned, the name {`difference predicates'} comes from the fact that they 
express a relation between the values of different functional expressions. 
For instance, 
the atom \mbox{\tt diff1(X,N2a,N2)}
expresses the relation 
between the value {\tt N2a} of the expression {\tt (count X S1)} and the value {\tt N2} of
 the expression  {\tt (count~X~(ins~X~S1))}. 

Actually, one can also show that there is a tight 
correspondence between the definition of the difference predicates 
and the lemmata which are needed for making the inductive proofs of 
the properties to
be shown. In particular, in the case of property {\it IS\_Perm}, if we replace 
in clause~{\tt15} defining {\tt diff1(X,N2a,N2)} its model 
 computed by Eldarica in the process of showing the satisfiability of 
the final set of clauses, that is, {\tt N2\,=\,N2a\,+\,1}~$\wedge$~{\tt N2a\,>=\,0},
from clause~{\tt 15} we get the following
lemma which is required for the proof of {\it IS\_Perm}
by induction on the list structure, namely, 

\smallskip
$\mathtt{\forall}$~{\tt X}, {\tt S1}, {\tt S}, 
{\tt N2}, {\tt N2a}. 
~{\tt ins(X,S1,S)} ~$\wedge$~ {\tt count(X,S,N2)} ~$\wedge$~ {\tt count(X,S1,N2a)} 
$~\rightarrow~$ 

\hspace*{33mm}({\tt N2\,=\,N2a\,+\,1} ~$\wedge$~ {\tt N2a\,>=\,0})

\smallskip
\noindent
Similarly to the above proof of  property~{\it IS\_Perm}, one can prove that program {\it InsertionSort} also satisfies the following property:

\smallskip
$\mathtt{\forall\, l,s.}$ {\tt (insertionSort l)$\,=\,$s}   
~~~$\rightarrow$~~~ {\tt(ordered s)}
\hfill $(\mathit{IS\_Orderedness})$\hspace{10mm}

\smallskip
\noindent
where the function {\tt ordered} from integer lists to booleans is defined 
as follows:

\vspace{-2mm}

\begin{verbatim}
let rec ordered l =
  match l with
  | Nil -> true
  | Cons(x,xs) -> match xs with 
                  | Nil -> true
                  | Cons(y,ys) -> x<=y & (ordered xs);;
\end{verbatim}

\vspace{-2mm}

\noindent
We start from the initial clauses~{\tt2}--{\tt9} together with the following clause which
translates the property~$\mathit{IS\_Orderedness}$:

\vspace{-1mm}

\begin{verbatim}
  false :- B=false, insertionSort(L,S), ordered(S,B).
\end{verbatim}

\vspace{-1mm}
\noindent
and after a transformation similar to the one we have described for property 
{\it IS\_Perm}, we get the following final set of clauses:

\vspace{-1mm}

\begin{verbatim}
  false :- B=false, new1(B).
  new1(true).
  new1(B) :- new1(B1), diff(X,B,B1).
  diff(I,true,true).
  diff(I,B,B).
  diff(I,false,false).
  diff(I,true,true).
  diff(I,B,B1) :- new2(I,Y1,Y,B,B1).
  new2(I,Y1,Y,B,B) :- I=<Y1, I=Y.
  new2(I,Y1,Y,false,false) :- I>Y, Y1=Y.
  new2(I,Y1,Y,true,true) :- I>Y, Y1=Y.
  new2(I,Y1,Y,B,B1) :- I>Y, Y=<Y2, Y=<Y3, Y1=Y, new2(I,Y3,Y2,B,B1).
\end{verbatim}

\vspace{-1mm}

\noindent
whose satisfiability can be shown by Eldarica. The model that Eldarica finds is the
following:

\begin{verbatim}
  new1(A) :- (A = true).
  diff(A,B,C) :- (\+((C = true)); (B = true)).
  new2(A,B,C,D,E) :- (\+((E = true)); (D = true)).
\end{verbatim}

\noindent
Having shown the properties {\it IS\_Perm} and {\it IS\_Orderedness} 
the automatic proof of 
partial correctness of program {\it InsertionSort} is completed.

\medskip

We have also proved various other properties of linearly recursive 
sorting programs using the
techniques we have presented above based on the application of the
Elimination Algorithm and the introduction of difference predicates.

In particular, we have shown that {\it InsertionSort} satisfies the following properties:

\smallskip
\noindent
\makebox[6mm][l]{(i)}$\forall\, {\mathtt{l}},{\mathtt{s}}.~~{\mathtt{(insertionSort~l)~=~s
~\rightarrow~  (length~l)~=~(length~s)}}$\hfill\makebox[24mm][l]{$(\mathit{IS\_Length})$}

\smallskip
\noindent
\makebox[6mm][l]{(ii)}$\forall\, {\mathtt{l}},{\mathtt{s}}.~~{\mathtt{(insertionSort~l)~=~s
		~\rightarrow~  (sumlist~l)~=~(sumlist~s)}}$\hfill\makebox[24mm][l]{$(\mathit{IS\_Sum})$}

\smallskip
\noindent 
where  {\tt length l}~computes the numbers of elements in the 
list~{\tt l}, and {\tt sumlist l}~computes the sum of the elements in the 
integer list~{\tt l}.

Moreover, for the following {\it SelectionSort}  program:

\vspace{-1mm}

\begin{verbatim}
let rec min (Cons(x,xs)) =
   match xs with
   | Nil -> x
   | Cons(y,ys) -> if x<y then min (Cons(x,ys)) else min (Cons(y,ys));;
let rec delete m (Cons(x,xs)) =
   if m=x then xs else Cons(x,delete m xs);;
let rec selectionSort l =
   match l with
   | Nil -> Nil
   | Cons(x,xs) -> let m = min (Cons(x,xs)) in
                      (Cons(m,selectionSort(delete m (Cons(x,xs)))));;
\end{verbatim}

\vspace{-1mm}

\noindent
we have shown the following three properties:

\smallskip
\noindent
\makebox[8mm][l]{(i)}\makebox[14mm][l]{$\mathtt{\forall\, l,s,x.}$}{\tt (selectionSort~l)$\,=\,$s}   
~~~$\rightarrow$~~~ {\tt(count x l)$\,=\,$(count x s)}
\hfill\makebox[29mm][l]{$(\mathit{SS\_Perm})$}

\smallskip
\noindent
\makebox[8mm][l]{(ii)}\makebox[14mm][l]{$\mathtt{\forall\, l,s.}$}{\tt (selectionSort~l)$\,=\,$s}   
~~~$\rightarrow$~~~ {\tt(ordered s)}
\hfill\makebox[29mm][l]{$(\mathit{SS\_Orderedness})$}

\smallskip
\noindent
\makebox[8mm][l]{(iii)}\makebox[14mm][l]{$\mathtt{\forall\, l,s.}$}{\tt 
(selectionSort~l)$\,=\,$s}   
~~~$\rightarrow$~~~ {\tt(length~l)$\,=\,$(length~s)}
\hfill\makebox[29mm][l]{$(\mathit{SS\_Length})$}

\section{Verification of Non-Linearly Recursive Sorting Algorithms}
\label{NonLinearRecursionVerification}
Now, we consider two popular non-linearly recursive sorting algorithms: 
{\em QuickSort} and {\em MergeSort}, and we show how our
transformation-based method allows us to verify some properties of those
algorithms.

Let us first consider the following {\em QuickSort} program which implements
a simplified version of the QuickSort algorithm: 
\begin{verbatim}
let rec partition x l = 
  match l with 
  | Nil -> (Nil,Nil) 
  | Cons(y,ys) -> let (l1,l2) = partition x ys in 
                  if x>=y then (Cons(y,l1),l2) else (l1,Cons(y,l2));;
let rec append l ys = 
  match l with 
  | Nil -> ys 
  | Cons(x,xs) -> Cons(x, append xs ys);;
let rec quickSort l = 
  match l with
  | Nil -> Nil
  | Cons(x,xs) -> let (ys,zs) = partition x xs in 
                  append (quickSort ys) (Cons(x,(quickSort zs)));;
\end{verbatim}

\noindent
where, given an integer {\tt x} and an integer list {\tt l}, {\tt partition x l}
returns a pair {\tt (l1,l2)} of lists such that their concatenation is {\tt l} 
itself and all elements of {\tt l1} are all the elements of {\tt l}
not larger than {\tt x}. Functions {\tt append} and {\tt quickSort} 
behave as expected.

Similarly to the examples of Section~\ref{LinearRecursionVerification} we want to prove
the following property of 
{\it QuickSort\,}: 

\smallskip
$\mathtt{\forall\, l,s,x.}$ {\tt (quickSort l)$=$s}   
~~~$\rightarrow$~~~ {\tt(count x l)$=$(count x s)}
\hfill $(\mathit{QS\_Perm})$\hspace{10mm}

\smallskip
\noindent
which states that the list {\tt s} returned by {\tt quickSort l} 
is a permutation of the list {\tt l}.

Program {\it QuickSort} and property $\mathit{QS\_Perm}$ are translated into the following constrained Horn clauses:

\vspace{-1mm}

\begin{verbatim}
1. false :- M=\=N, quickSort(L,S), count(X,L,M), count(X,S,N).
2. partition(X,[],[],[]).
3. partition(X,[Y|Ys],[Y|L1],L2) :- Y=<X, partition(X,Ys,L1,L2). 
4. partition(X,[Y|Ys],L1,[Y|L2]) :- X<Y, partition(X,Ys,L1,L2). 
5. append([],Ys,Ys).
6. append([X|Xs],Ys,[X|Zs]) :- append(Xs,Ys,Zs).
7. quickSort([],[]). 
8. quickSort([X|Xs],S) :- partition(X,Xs,Ys,Zs), 
      quickSort(Ys,S1), quickSort(Zs,S2), append(S1,[X|S2],S).
\end{verbatim}

\vspace{-1mm}

\noindent
The clauses for {\tt count} are as in Section~\ref{LinearRecursionVerification}.
Now, we transform the set made out of clauses~{\tt1}--{\tt8}, together with 
the clauses for {\tt count},
into a new set of clauses without lists.
We start off by introducing a new predicate 
{\tt new1} by the following clause (note that the arguments of {\tt new1} are the
non-list variables occurring in the body):

\vspace{-1mm}

\begin{verbatim}
9. new1(X,M,N) :- quickSort(L,S), count(X,L,M), count(X,S,N).
\end{verbatim}

\vspace{-1mm}

\noindent
By folding clause {\tt1}, we derive a new clause without occurrences of lists:

\vspace{-1mm}

\begin{verbatim}
10. false :- M=\=N, new1(X,M,N).
\end{verbatim}

\vspace{-1mm}

\noindent
We proceed by unfolding clause~{\tt9} and we derive the following clauses,
where we have rearranged the order of the atoms in the bodies to 
simplify the explanation of the subsequent steps:

\vspace{-1mm}

\begin{verbatim}
11. new1(X,0,0).
12. new1(X,M,N) :- M=K+1, X=Y,  
       quickSort(L1,S1), quickSort(L2,S2), count(X,L,K), partition(Y,L,L1,L2), 
       append(S1,[Y|S2],S), count(X,S,N).
13. new1(X,M,N) :- X=\=Y, 
       quickSort(L1,S1), quickSort(L2,S2), count(X,L,M), partition(Y,L,L1,L2), 
       append(S1,[Y|S2],S), count(X,S,N).
\end{verbatim}

\vspace{-1mm}

\noindent 
It is impossible to fold clauses~{\tt12} and~{\tt13}
using clause~{\tt9}, and thus, similarly to the {\it InsertSort} 
example in Section~\ref{LinearRecursionVerification},
we apply the {difference predicate} technique.
However, the presence of the two recursive calls to {\tt quickSort}, 
makes things a bit more involved in this case. 
We show the derivation relative to clause~{\tt 12}. The one relative to 
clause~{\tt13}
is similar and we omit it.

\smallskip
\noindent
$\bullet$ {\it Step} 1. {\it Embed}. 
The atoms {\tt quickSort(L1,S1), count(X,L,K), count(X,S,N)} in the body of 
clause~{\tt12} are 
individually variants of atoms in the body of clause {\tt9}. However, 
their conjunction is {\it not} a 
variant of the body of clause~{\tt9}, and thus we cannot fold clause~{\tt 12}
using clause~{\tt9}.
Similarly, {\tt quickSort(L2,S2), count(X,L,K), count(X,S,N)} is not a 
variant of the body of clause~{\tt9}.

\smallskip
\noindent
$\bullet$ {\it Step} 2. {\it Rename}. In order to fold the two {\tt quickSort} atoms
in the body of clause {\tt12} we consider two renamed apart variants of clause {\tt 9}. We get:

	\begin{verbatim}
9a. new1(Xa,Ma,Na) :- quickSort(La,Sa), count(Xa,La,Ma), count(Xa,Sa,Na).
9b. new1(Xb,Mb,Nb) :- quickSort(Lb,Sb), count(Xb,Lb,Mb), count(Xb,Sb,Nb).
	\end{verbatim}

\smallskip
\noindent
$\bullet$ {\it Step} 3. {\it Match}. We simultaneously match clauses~{\tt9a} 
and~{\tt9b}  against clause~{\tt12} to be folded. We 
get the substitution 
$\sigma= \{{\mathtt{La/L1,}}$ ${\mathtt{Sa/S1,}}$ 
${\mathtt{Xa/X,}}$ ~ ${\mathtt{Lb/L2,}}$ ${\mathtt{Sb/S2,}}$ ${\mathtt{Xb/X}}\}$.
This substitution cannot be extended so to include more bindings obtained 
by pairs of {\tt count} atoms, because: (i) we cannot add the 
bindings $\mathtt{La/L,}$
${\mathtt{Sa/S,}}$ ${\mathtt{Lb/L,}}$ ${\mathtt{Sb/S,}}$ which are inconsistent with 
the bindings for the same variables already in $\sigma$, and
(ii) we cannot add bindings for the output variables
$\mathtt{Ma}$, $\mathtt{Mb}$, $\mathtt{Na}$, $\mathtt{Na}$,
because {we stipulate that} the output variables of two predicates may match 
only if all their input variables match~\footnote{The input 
and output variables of the predicate 
{\tt count} are defined as expected, if one considers the function
{\tt count} of the functional program it comes from.}. 
For instance, as already said, the matching of the input variables
$\mathtt{La}$ and $\mathtt{L}$ of the atoms $\mathtt{count(Xa,La,Ma)}$ and 
$\mathtt{count(X,L,M)}$ fails, and hence we cannot add the binding 
$\mathtt{Ma/M}$ for their output variables.

By applying $\sigma$ we get the following two instances of clauses {\tt 9a} and {\tt 9b}:

\smallskip   
\noindent
${\mathtt{9am.~~~new1(X,Ma,Na)~:\mbox{\tt -}~quickSort(L1,S1), ~||~ 
count(X,L1,Ma),
   ~count(X,S1,Na).}}$ 
      
\noindent
${\mathtt{9bm.~~~new1(X,Mb,Nb)~:\mbox{\tt -}~quickSort(L2,S2), ~||~ count(X,L2,Mb),
   ~count(X,S2,Nb).}}$
   
\smallskip   
\noindent
These clauses are the two clauses we will actually use for folding clause~{\tt12}.
Thus, the only atoms that are made equal by $\sigma$ are the two {\tt quickSort}
atoms in the bodies of clauses {\tt9am} and {\tt9bm} (which
we place to the left of the marker~\bbar), and 
the left and right {\tt quickSort} atoms in the body of clause~{\tt 12}, respectively.
As in clause~{\tt 10m} of 
Section~\ref{LinearRecursionVerification}, 
the marker~\bbar~separates the matching
conjunction to its left from the mismatching conjunction to its right.

The mismatching conjunctions of clauses~{\tt 9am} and~{\tt9bm}
we will use for folding (which
we place to the right of the markers~\bbar)
are made out of the atoms {\tt count(X,L1,Ma)}, {\tt count(X,S1,Na)}, 
and {\tt count(X,L2,Mb)}, {\tt count(X,S2,Nb)}, respectively, while 
the mismatching conjunction of 
clause~{\tt12} to be folded is made out
of the following atoms:

\smallskip
{\tt count(X,L,K)}, ~~{\tt partition(Y,L,L1,L2)}, 
~~{\tt append(S1,[Y|S2],S)}, ~~{\tt count(X,S,N)}. \hfill{(*)}~~

\smallskip
\noindent
$\bullet$ {\it Step} 4. {\it Introduce Difference Predicates}. 
Now, in order to fold clause {\tt12} using the two clauses 
{\tt 9am} and {\tt9bm}, we
need to replace the mismatching conjunction of clause~{\tt12}
by the mismatching conjunctions of clauses~{\tt 9am} and~{\tt9bm}.
This replacement 
can be done at the expense of adding to the body of clause~{\tt 12} two 
new atoms with {\it difference predicates}. 
As already remarked, this atom addition is required for preserving 
satisfiability of the derived clauses. 
In our case the difference predicates we introduce, called {\tt diff1} and {\tt
diff2}, are 
defined as follows: 

\smallskip
\noindent
{\tt 14. diff1(X,Y,K,Ma,Mb) :- X=Y, count(X,L,K), partition(Y,L,L1,L2),} ~~$\tdagger$\nopagebreak

\hspace*{48mm}{\tt count(X,L1,Ma), count(X,L2,Mb).}\nopagebreak

\noindent
{\tt 15. diff2(X,Y,N,Na,Nb) :- X=Y, append(S1,[Y|S2],S), count(X,S,N),} ~~$\tdagger$\nopagebreak

\hspace*{48mm}{\tt count(X,S1,Na), count(X,S2,Nb).}

\smallskip

\noindent
{These two clauses} are constructed as we now specify.
(i)~We first consider the set of atoms occurring in the 
mismatching conjunction of the clause to be folded (in our case the atoms~(*) of 
clause~{\tt12}) and in the mismatching conjunctions of the clauses we will use 
for folding (in our case 
the atoms to the right of~\bbar~in clauses~{\tt9am} and {\tt9bm}),
and then we partition that set into two subsets 
by keeping together only the atoms that share common list variables.
These two subsets constitute the bodies of the clauses~{\tt14} and~{\tt15}.
(ii)~We add the constraint {\tt X=Y} 
to the bodies of clause~{\tt14} and~{\tt15} to define more specific predicates as needed in the 
clause~{\tt12} to be folded, where the same constraint occurs.
(iii)~We specify the arguments of {\tt  diff1} and {\tt diff2}
to be the non-list variables of the corresponding bodies (obviously, the order
of the arguments is insignificant).

The markers~$\tdagger$ in clauses~{\tt14} and~{\tt15} separate the atoms 
to be removed from the body of clause {\tt12} (those to the left)
from the atoms to be added to the body of clause {\tt12} (those to the right)
so that the folding steps can then be performed.

The reader may note that the difference predicates express
the relations between the non-list output variables of the atoms 
that are added and the 
non-list output variables of the atoms that are removed. In particular, 
for clause~{\tt14}, the atom {\tt diff1(X,Y,K,Ma,Mb)} expresses
the relation between the values of the output variables 
{\tt K}, {\tt Ma}, and {\tt Mb} of the {\tt count} atoms,
for all values of the input variables {\tt X} and {\tt Y} of the {\tt count} atoms.
(The output variables {\tt L1} and {\tt L2} of the removed atom {\tt partition(Y,L,L1,L2)} do not play any role because they are list 
variables.)

Analogously, for clause~{\tt15}, the atom \mbox{\tt diff2(X,Y,N,Na,Nb)}
expresses
the relation between the values of the output variables
{\tt N}, {\tt Na}, and {\tt Nb} of the {\tt count} atoms,
for all values of the input variables {\tt X} and {\tt Y} of the {\tt count} atoms.

Note that we could have introduced, instead of two difference predicates,
a single difference predicate by a clause whose body was the conjunction of the atoms in the
bodies of clauses~{\tt 14} and~{\tt 15}. Thus, that conjunction 
would have been made out of exactly the mismatching
conjunctions of the clause to be folded (clause~{\tt12}) and the clauses used for 
folding (clauses~{\tt9am} and~{\tt9bm}), as in the case of the {\it InsertSort}
example in Section~\ref{LinearRecursionVerification}.
Also this single, more complex difference predicate would have 
allowed the elimination of lists. 

\smallskip
\noindent
$\bullet$ {\it Step} 5. {\it Replace}. In clause~{\tt12} we replace the atoms
to the left of~$\tdagger$ in the bodies of clauses~{\tt 14} and~{\tt 15},
by the atoms to the right of~$\tdagger$, and we also add the heads of 
clauses~{\tt 14} and~{\tt 15}. We get (after a reordering of the atoms):

\vspace{-1mm}

\begin{verbatim}
16. new1(X,M,N) :- M=K+1, X=Y, diff1(X,Y,K,Ma,Mb),
         quickSort(L1,S1), count(X,L1,Ma), count(X,S1,Na), 
         quickSort(L2,S2), count(X,L2,Mb), count(X,S2,Nb), diff2(X,Y,N,Na,Nb). 
\end{verbatim}

\vspace{-1mm}

\noindent
It can be shown that, if the values of {\tt K} and {\tt N} which are 
arguments of the difference predicates {\tt diff1} and {\tt diff2},
are functions of
({\tt X,Y,Ma,Mb}) and ({\tt X,Y,Na,Nb}), respectively, then 
the clauses before this replacement are satisfiable {\it if and only if\/} the clauses after 
the replacement are satisfiable.
The proof of this fact is outside the scope of this paper.

\smallskip
\noindent
$\bullet$ {\it Step} 6. {\it Fold}. Now, we are able to fold twice clause~{\tt 16} 
using clauses~{\tt 9am} and {\tt9bm}, and we derive the following clause, 
where all arguments are integer variables:

\vspace{-1mm}

\begin{verbatim}
17. new1(X,M,N) :- M=K+1, X=Y, diff1(X,Y,K,Ma,Mb), new1(X,Ma,Na), 
                   new1(X,Mb,Nb), diff2(X,Y,N,Na,Nb). 
\end{verbatim}

\noindent
{Now, as the reader may check, if we start from clause {\tt 13} and we perform in a 
similar way the six steps  
we have performed starting from clause {\tt 12} for deriving clause~{\tt17}, we get the following clause:}

\vspace{-1mm}

\begin{verbatim}
18. new1(X,M,N) :- X=\=Y, diff3(X,Y,M,Ma,Mb), new1(X,Ma,Na), 
                   new1(X,Mb,Nb), diff4(X,Y,N,Na,Nb).
\end{verbatim}

\vspace{-1mm}

\noindent
where {\tt diff3} and {\tt diff4} are difference predicates defined by the following clauses: 

\vspace{-1mm}

\begin{verbatim}
19. diff3(X,Y,M,Ma,Mb) :- X=\=Y, count(X,L,M), partition(Y,L,L1,L2),
                          count(X,L1,Ma), count(X,L2,Mb).
20. diff4(X,Y,N,Na,Nb) :- X=\=Y, append(S1,[Y|S2],S), count(X,S,N),
                          count(X,S1,Na), count(X,S2,Nb).
\end{verbatim}

\vspace{-1mm}

At this point we can derive clauses without lists for {\tt diff1}--{\tt diff4}
by applying the Elimination Algorithm~\cite{De&18a}. Indeed, by introducing some new 
predicates, namely {\tt new2}, {\tt new3}, and {\tt new4}, we get:

\vspace{-1mm}

\begin{verbatim}
21. diff1(X,Y,0,0,0)   :- X=Y.
22. diff1(X,Y,M,Ma,Mb) :- X=Y, Ma=Ma1+1, M=M1+1, diff1(X,Y,M1,Ma1,Mb).
23. diff2(X,Y,N,0,Nb)  :- X=Y, N=Nb+1, new2(X,Nb).
24. diff2(X,Y,N,Na,Nb) :- X=Y, Na=Na1+1, N=N1+1, diff2(X,Y,N1,Na1,Nb).
25. diff3(X,Y,M,Ma,Mb) :- X<Y, new3(X,Y,M,Ma,Mb).
26. diff3(X,Y,M,Ma,Mb) :- X>Y, new4(X,Y,M,Ma,Mb).
27. diff4(X,Y,N,0,N)   :- X=\=Y, new2(X,N).
28. diff4(X,Y,N,Na,Nb) :- X=\=Y, Na=Na1+1, N=N1+1, diff4(X,Y,N1,Na1,Nb).
29. new2(X,0).
30. new2(X,N) :- N=N1+1, new2(X,N1).
31. new3(X,Y,0,0,0)   :- X<Y.
32. new3(X,Y,M,Ma,Mb) :- X<Y, M=M1+1, Ma=Ma1+1, new3(X,Y,M1,Ma1,Mb).
33. new4(X,Y,0,0,0)   :- X>Y.
34. new4(X,Y,M,Ma,Mb) :- X>Y, M=M1+1, Mb=Mb1+1, new4(X,Y,M1,Ma,Mb1).
\end{verbatim}

\vspace{-1mm}

The final set of clauses without list arguments consists of clauses
{\tt 10}, {\tt11}, {\tt17}, {\tt18}, {\tt21}--{\tt34}. 
Eldarica is able to prove the satisfiability of this set of clauses
by computing a model, which is expressible in LIA~(see Appendix).
Thus, we have proved that property {\it QS\_Perm} holds for the {\it QuickSort} program.

\smallskip
For {\it QuickSort} we have also proved 
the following arithmetic property by using the same technique shown above:

\smallskip
$\forall\, {\mathtt{l}},{\mathtt{s}}.~~{\mathtt{(quickSort~l)~=~s
 ~\rightarrow~ (sumlist~l)~=~(sumlist~s)}}$
 \hfill\makebox[24mm][l]{$(\mathit{QS\_Sum})$}

\smallskip

\noindent 
Now as a second example of non-linearly recursive sorting programs, let us now consider the 
following {\it MergeSort} program:

\vspace{-1mm}

\begin{verbatim}
let rec takedrop d l = 
   match l with
   | Nil -> (Nil,Nil)
   | Cons(x,xs) -> if d=0 then (Nil,l) else 
                   let (a,b) = takedrop (d-1) xs in (Cons(x,a),b);;
let rec split l = 
   let d = (length l)/2 in takedrop d l;; 
\end{verbatim}

\vspace{-1mm}

\begin{verbatim}
let rec merge l1 l2 = 
   match l1 with 
   | Nil -> l2 
   | Cons(x,xs) -> match l2 with
                   | Nil -> l1
                   | Cons(y,ys) -> if x<y then Cons(x,merge xs l2) else 
                           Cons(y, merge l1 ys);;
\end{verbatim}

\vspace{-1mm}

\begin{verbatim}
let rec mergeSort l = 
   match l with
   | Nil -> Nil
   | Cons(x,xs) -> 
         match xs with
         | Nil -> Cons(x,Nil)
         | Cons(y,ys) -> let (fst,snd) = split l in
                            merge (mergeSort fst) (mergeSort snd);; 
\end{verbatim}

\vspace{-1mm}

\noindent 
where, given an integer {\tt d} 
and a list {\tt l}, {\tt takedrop d l} returns a pair {\tt (fst,snd)} of lists
such that: (i)~{\tt fst} is a list made out of the first {\tt d} elements of {\tt l}, 
and (ii)~{\tt snd} 
is a list made out of the remaining elements of~{\tt l}. The other functions
of {\it MergeSort}
behave as expected. Note that the value of {\tt d} in the function call 
{\tt split~l} is obtained by integer division, and thus {\tt d} 
is the largest integer smaller or equal
to the length of {\tt l} divided by~2.

For {\it MergeSort} we have proved
the following property:

\smallskip
$\forall\, {\mathtt{l}},{\mathtt{s}}.~~{\mathtt{(mergeSort~l)~=~s
 ~\rightarrow~ (sumlist~l)~=~(sumlist~s)}}$\hfill\makebox[24mm][l]{$(\mathit{MS\_Sum})$}

\smallskip
\noindent
The CHC translations, the transformed CHCs, and the models computed by Eldarica for
all verification problems considered in this section and in the previous one
are presented in the Appendix.

{We think that our method
	based on the introduction of difference predicates works for
	many other verification problems with a structure similar to
	the ones presented in this paper. 
	We leave for future work the task of making a thorough experimental evaluation.
}

\section{Concluding Remarks}
\label{ConcludingRemarks}
Let us briefly discuss how 
the transformation-based verification technique presented 
in this paper, through some sorting examples,
can be applied to a larger class of verification problems.
We will consider in particular: (A)~the correctness issue of that technique, 
and (B)~the mechanization of that technique.

Point (A).~The main hypothesis needed to show correctness 
of our  technique is that the predicates occurring in the
initial set of clauses define total functional relations.
This property is guaranteed by construction in the case when those predicates 
are the CHC translation of functional programs that terminate for all inputs. 
One more hypothesis that we need is the functionality of the difference predicates 
introduced during the transformation process. This functionality requirement can be checked 
in the model computed by the CHC solver, which is expressed as
a set of LIA and/or Bool constraints. 
{Note, however, that even if functionality of the difference
predicates does not hold, it is the case that 
the satisfiability of the derived clauses implies the 
satisfiability of the initial clauses, and in many practical 
verification problems this implication is all that is required.}

Point (B).~In order to mechanize our transformation technique,
we need to extend the Elimination Algorithm~\cite{De&18a} with a suitable
automated mechanism for introducing difference predicates.
This mechanism can be based on the six steps
shown in our examples of Sections~\ref{LinearRecursionVerification} and~\ref{NonLinearRecursionVerification}.
Among these, Step 3 (Matching) seems to be the most critical one,
as many different matching substitutions could be computed for the same pair
of clauses. However, once a matching has been chosen among the possible
ones, the subsequent steps, and in particular the introduction of
the difference predicates (Step 4), is straightforward.
More sophisticated mechanisms may take into account the constraints
occurring in the clauses, and may apply widening techniques which have been 
considered in other transformation methods~\cite{De&14c,Ka&16}.
At the time of writing we have made some initial steps towards 
an implementation of the extended Elimination Algorithm that introduces
difference predicates by using the VeriMAP transformation and 
verification system~\cite{De&14b}.

{Various techniques for mechanizing induction over 
data structures have been proposed in the literature 
(see, for instance, the {\em rippling} technique~\cite{Bun01}),
but in the \mbox{two-phase} approach to program verification 
we apply in this paper
we need not to take advantage of them.
Indeed, in the first phase of our approach 
we transform using fold/unfold rules a set of clauses 
into a new, equisatisfiable set of clauses where the 
data structures have been eliminated, and thus no induction over them
is required. Then,
in the second phase of our approach, we do not need
induction either, because satisfiability 
of the new set of clauses 
is shown by the use of CHC solvers acting over 
Linear Integer Arithmetic and Boolean constraints only.}


To summarize, 
this paper presents ongoing work which follows a very general approach
to program verification based on constrained Horn clauses.
As also shown in the examples we have presented, the reduction of a program
verification problem to a CHC satisfiability problem can often be 
obtained by a straightforward translation. However, proving the satisfiability
of the clauses obtained by that translation is, in many cases, a much harder
task. In a series of papers~\cite{De&14c,De&15c,De&17b,De&18a,De&18c,Ka&16} 
it has been shown that by combining various 
transformation techniques, such as {\em Specialization} and 
{\em Predicate Pairing},
we can derive equisatisfiable sets of clauses for which the efficacy of the
CHC solvers is significantly improved. This approach avoids the 
burden of implementing very sophisticated solving strategies depending 
on the class of satisfiability problems {one is required to solved.} 
In particular, in the class of problems considered in this paper,
consisting in checking the satisfiability of clauses over inductively 
defined data structures,
we can avoid to implement {\it ad hoc} strategies to deal with induction proofs.
{In this paper we have also investigated the use of a novel transformation
technique~\cite{De&19a} based 
on the introduction of the so-called difference predicates and their relation with
the lemmata that are often necessary in the inductive proofs of the properties of interest.}

To the best of our knowledge, the only other approach that makes use of
transformations of CHCs to remove inductively defined data structures is
the one based on the notion of {\em CHC product}~\cite{MoF17}.
A CHC product is an operation that composes pairs of sets of clauses that have
a similar recursive structure and, in some cases, it derives
more compact sets of clauses with respect to those 
obtained by fold/unfold transformations.
On the other hand, however, fold/unfold transformations 
are more general than CHC products,
and {often} allow easier proofs of correctness. 
Moreover, no mechanism for lemma generation is provided when 
computing CHC products.

We leave it for future work to experiment on significant benchmarks 
and test whether our approach pays off in practice.

\section{Acknowledgments}
\label{Acknow}
This work was supported by Gruppo Nazionale per il Calcolo Scientifico (GNCS-INdAM).\\
Emanuele De Angelis, Fabio Fioravanti, and Alberto Pettorossi are research associates at
CNR-IASI, Rome, Italy.


\newpage
\section{Appendix}

\noindent
In this Appendix we present the proofs
of some properties of various functional 
programs implementing sorting algorithms, including those presented in the 
previous sections. 
For every proof relative to a 
$\langle$program, property$\rangle$ pair
we present:

\noindent
(1)~the set of 
constrained Horn clauses produced by the translation of the 
functional sorting program and the property
to be proved, 

\noindent
(2)~the set of constrained Horn 
clauses derived after the transformation based on the Elimination
Algorithm~\cite{De&18a} extended by the technique for
introducing difference predicates presented in this paper, and 

\noindent
(3)~the model of the 
constrained Horn clauses derived at Point~(2) which has been computed by  the CHC solver Eldarica.


Eldarica has been called by the command 
`{\tt \$eldarica -horn -hsmt -sol} \hspace{2mm}{\it filename}',
where \mbox{\it filename} {is the file containing the SMT-LIB translation of the CHCs for which a model is sought.}

Note that Eldarica has not been able to construct a model {for any} of the 
sets of constrained Horn 
clauses of Point~(1) (that is, before the application of the Elimination Algorithm 
extended by the introduction of difference predicates) within the timeout of 120 
second.
All proofs  have
been conducted on a personal computer  Intel Core i7-8550U CPU, 1.80GHz X 4 
cores (8 threads) with 16 GB of RAM.  

\smallskip
\noindent
The $\langle$program, property$\rangle$ pairs we have considered in this appendix are:

\smallskip
\noindent
\makebox[65mm][l]{\makebox[5mm][r]{1.~} $\langle${\it InsertionSort}, ~{\it IS\_Perm}$\rangle$}
\makebox[5mm][r]{2.~} $\langle${\it InsertionSort}, ~{\it IS\_Orderedness}$\rangle$

\smallskip
\noindent
\makebox[65mm][l]{\makebox[5mm][r]{3.~} $\langle${\it InsertionSort}, ~{\it IS\_Length}$\rangle$}
\makebox[5mm][r]{4.~} $\langle${\it InsertionSort}, ~{\it IS\_Sum}$\rangle$

\smallskip
\noindent
\makebox[65mm][l]{\makebox[5mm][r]{5.~} $\langle${\it SelectionSort}, ~{\it SS\_Perm}$\rangle$}
\makebox[5mm][r]{6.~} $\langle${\it SelectionSort}, ~{\it SS\_Orderedness}$\rangle$

\smallskip
\noindent
\makebox[65mm][l]{\makebox[5mm][r]{7.~} $\langle${\it SelectionSort}, ~{\it SS\_Length}$\rangle$}
\makebox[5mm][r]{8.~} $\langle${\it QuickSort}, ~{\it QS\_Perm}$\rangle$

\smallskip
\noindent
\makebox[65mm][l]{\makebox[5mm][r]{9.~} $\langle${\it QuickSort}, ~{\it QS\_Sum}$\rangle$}
\makebox[5mm][r]{10.~} $\langle${\it MergeSort}, ~{\it MS\_Sum}$\rangle$


\subsection{{\it {\rm Program} InsertionSort} {\rm and property} {\it IS\_Perm.}}

\subsubsection{{\large{\rm Constrained Horn clauses obtained by translating 
{\it InsertionSort} and {\it IS\_Perm}.}}}

\begin{verbatim}
:- pred ins(int,list(int),list(int)).
:- pred insertionSort(list(int),list(int)).
:- pred count(int,list(int),int).

false :- C1=\=C2, insertionSort(L,S), count(X,L,C1), count(X,S,C2).
ins(I,[],[I]).
ins(I,[X|Xs],[I,X|Xs]) :-  I=<X.
ins(I,[X|Xs],[X|Ys]) :-  I>X, ins(I,Xs,Ys).
insertionSort([],[]).
insertionSort([X|Xs],S) :- insertionSort(Xs,S1), ins(X,S1,S).
count(X,[],0).
count(X,[H|T],N) :- X=H, N=M+1, count(X,T,M).
count(X,[H|T],N) :- X=\=H, count(X,T,N).
\end{verbatim}

\subsubsection{{\large{\rm Transformed constrained Horn clauses.}}}

\begin{verbatim}
:- pred new1(int,int,int).
:- pred new2(int,int).
:- pred new3(int,int,int).
:- pred new4(int,int,int,int).
:- pred new5(int,int).
:- pred new6(int,int,int,int).
:- pred diff1(int,int,int).
:- pred diff2(int,int,int,int).

false :- N1=\=N2, new1(X,N1,N2).
new1(X,0,0).
new1(X,N1,N2) :- N1=N+1, new1(X,N,N2a), diff1(X,N2a,N2).
new1(X,N1,N2) :- X=\=Y, new1(Y,N1,N2b), diff2(X,Y,N2b,N2).
diff1(X,0,N2)  :- N2=N1+1, new2(X,N1).
diff1(X,N1,N2) :- N2=M2+1, N1=M1+1, new3(X,M2,M1).
diff1(X,N1,N2) :- X=<Y, N2=N+1, X=\=Y, new4(X,Y,N,N1).
diff2(X,Y,0,0) :- Y=\=X.
diff2(X,Y,M,N) :- X=<Y, Y=\=X, M=K+1, new3(Y,N,K).
diff2(X,Y,M,N) :- X=<Z, Y=\=X, Y=\=Z, N=M, new5(Y,N).
diff2(X,Y,M,N) :- X>Y, N=H+1, M=K+1, diff2(X,Y,K,H).
new2(X,0).
new3(X,N1,N)  :- N1=N+1, new5(X,N).
new4(X,Y,N,N) :- X=<Y, X=\=Y, new5(X,N).
new5(X,0).
new5(X,N1) :- N1=N+1, new5(X,N).
\end{verbatim}

\subsubsection{{\large{\rm Model of the transformed constrained Horn clauses.}}}

\begin{verbatim}
diff1(A,B,C) :- (((B - C) = -1), (B >= 0)).
diff2(A,B,C,D) :- ((C = D), (C >= 0)).
new1(A,B,C) :- ((B = C), (B >= 0)).
new2(A,B) :- (B = 0).
new3(A,B,C) :- (((C - B) = -1), (B >= 1)).
new4(A,B,C,D) :- ((D = C), ((C >= 0), ((B - A) >= 1))).
new5(A,B) :- (B >= 0).
\end{verbatim}

\subsection{{\it {\rm Program} InsertionSort} {\rm and property} 
{\it IS\_Orderedness.}}

\subsubsection{{\large{\rm Constrained Horn clauses obtained by translating 
{\it InsertionSort} and {\it IS\_Orderedness}.}}}

\begin{verbatim}
:- pred ins(int,list(int),list(int)).
:- pred insertionSort(list(int),list(int)).
:- pred ordered(list(int),bool).

false :- insertionSort(L,S), ordered(S,false).
ins(I,[],[I]).
ins(I,[X|Xs],[I, X| Xs]) :- I=<X.
ins(I,[X|Xs],[X|Ys]) :- I>X, ins(I,Xs,Ys).
insertionSort([],[]).
insertionSort([X|Xs],S) :- insertionSort(Xs,S1), ins(X,S1,S).
ordered([],true).
ordered([X],true).
ordered([X,Y|T],false) :-X>Y.
ordered([X,Y|T],B) :-X=<Y, ordered([Y|T],B).
\end{verbatim}

\subsubsection{{\large{\rm Transformed constrained Horn clauses.}}}

\begin{verbatim}
:- pred diff(int,bool,bool).
:- pred new1(bool).
:- pred new2(int,int,int,bool,bool).

false :- new1(false).
new1(true).
new1(B) :- new1(B1), diff(X,B,B1).
diff(I,true,true).
diff(I,B,B).
diff(I,false,false).
diff(I,B,B1) :- new2(I,Y1,Y,B,B1).
new2(I,Y1,Y,B,B) :- I=<Y1, I=Y.
new2(I,Y1,Y,false,false) :- I>Y, Y1=Y.
new2(I,Y1,Y,true,true) :- I>Y, Y1=Y.
new2(I,Y1,Y,B,B1) :- I>Y, Y=<Y2, Y=<Y3, Y1=Y, new2(I,Y3,Y2,B,B1).
\end{verbatim}

\subsubsection{{\large{\rm Model of the transformed constrained Horn clauses.}}}

\begin{verbatim}
new1(A) :- (A = true).
diff(A,B,C) :- (\+((C = true)); (B = true)).
new2(A,B,C,D,E) :- (\+((E = true)); (D = true)).
\end{verbatim}

\subsection{{\it {\rm Program} InsertionSort} {\rm and property} 
{\it IS\_Length.}}

\subsubsection{{\large{\rm Constrained Horn clauses obtained by translating 
{\it InsertionSort} and {\it IS\_Length}.}}}

\begin{verbatim}
:- pred length(list(int),int).
:- pred delete(int,list(int),list(int)).
:- pred ins(int,list(int),list(int)).
:- pred insertionSort(list(int),list(int)).

false :- N1>=0, N2>=0, N1 =\= N2, 
    length(L,N1), insertionSort(L,S), length(S,N2).
length([],0).
length([X|Xs],N) :- N1>=0, N = N1+1, length(Xs,N1).
delete(X,[Y|T],T) :- X=Y.
delete(X,[Y|T],[Y|D]) :- X=\=Y, delete(X,T,D).
ins(I,[],[I]).
ins(I,[X|Xs],[I, X| Xs]) :- I=<X-1.
ins(I,[X|Xs],[X|Ys]) :- I>=X, ins(I,Xs,Ys).
insertionSort([],[]).
insertionSort([X|Xs],S) :- ins(X,S1,S), insertionSort(Xs,S1).
\end{verbatim}

\subsubsection{{\large{\rm Transformed constrained Horn clauses.}}}

\begin{verbatim}
:- pred diff(int,int,int).
:- pred new1(int,int).
:- pred new2(int).

false :- new1(N1,N2), N1 >= 0, N2 >= 0, N1 =\= N2.
new1(0,0).
new1(N1,N2) :- N11>=0, N1=N11+1, new1(N11,N3), diff(X,N2,N3).
diff(I,N2,N3) :- N2=1, N3=0.
diff(I,N2,N3) :- I=<X-1, N2=N3+1, N3=N5+1, new2(N5).
diff(I,N2,N3) :- I>=X, N2=N4+1, N3=N5+1, diff(I,N4,N5).
new2(0).
new2(N) :- N1=N+1, new2(N1).
\end{verbatim}

\subsubsection{{\large{\rm Model of the transformed constrained Horn clauses.}}}

\begin{verbatim}
diff(A,B,C) :- ((B - C) = 1).
new1(A,B) :- ((A = B), (A >= 0)).
new2(A) :- (A =< 0).
\end{verbatim}

\subsection{{\it {\rm Program} InsertionSort} {\rm and property} 
{\it IS\_Sum.}}

\subsubsection{{\large{\rm Constrained Horn clauses obtained by translating 
{\it InsertionSort} and {\it IS\_Sum}.}}}

\begin{verbatim}
:- pred sumlist(list(int),int).
:- pred ins(int,list(int),list(int)).
:- pred insertionSort(list(int),list(int)).

false :- M=\=N, sumlist(L,M), insertionSort(L,S), sumlist(S,N).
sumlist([],0).
sumlist([X|Xs],M) :- M=X+N, sumlist(Xs,N).
ins(I,[],[I]).
ins(I,[X|Xs],[I,X|Xs]) :- I=<X.
ins(I,[X|Xs],[X|Ys]) :- I>X, ins(I,Xs,Ys).
insertionSort([],[]).
insertionSort([X|Xs],S) :- insertionSort(Xs,SXs), ins(X,SXs,S).
\end{verbatim}

\subsubsection{{\large{\rm Transformed constrained Horn clauses.}}}

\begin{verbatim}
:- pred diff(int,int,int).
:- pred new1(int,int).
:- pred new2(int).

false :- M=\=N, new1(M,N).
new1(0,0).
new1(M1,N1) :- M1=H+M0, new1(M0,N0), diff(H,N0,N1).
diff(H,0,N1)  :- N1=H.
diff(H,N0,N1) :- H=<X, N0=X+N2, N1=H+N0, new2(N2).
diff(H,N0,N1) :- H>X,  N0=X+N2, N1=X+N3, diff(H,N2,N3).
new2(0).
new2(N) :- N=X+N1, new2(N1).
\end{verbatim}

\subsubsection{{\large{\rm Model of the transformed constrained Horn clauses.}}}

\begin{verbatim}
diff(A,B,C) :- ((A + (B - C)) = 0).
new1(A,B) :- (A = B).
new2(A) :- true.
\end{verbatim}

\subsection{{\rm Program} {\it SelectionSort} {\rm and property} 
{\it SS\_Perm.}}

\subsubsection{\large{\rm Constrained Horn clauses obtained by translating 
{\it SelectionSort} and {\it SS\_Perm}.}}

\begin{verbatim}
:- pred min(list(int),int).
:- pred delete(int,list(int),list(int)).
:- pred selectionSort(list(int),list(int)).
:- pred count(int,list(int),int).

false :- N1 =\= N2, count(X,L,N1), selectionSort(L,S), count(X,S,N2).
min([X],X).
min([X|T],M) :- X=<M1, M=X, min(T,M1).
min([X|T],M) :- X>M1, M=M1, min(T,M1).
delete(X,[Y|T],T) :- X=Y.
delete(X,[Y|T],[Y|D]) :- X=\=Y, delete(X,T,D).
selectionSort([],[]).
selectionSort(L,[M|T]) :- min(L,M), delete(M,L,L1), selectionSort(L1,T).
count(X,[],0).
count(X,[H|T],N) :- X=H, N=M+1, count(X,T,M).
count(X,[H|T],N) :- X=\=H, count(X,T,N).
\end{verbatim}

\subsubsection{{\large{\rm Transformed constrained Horn clauses.}}}

\begin{verbatim}
:- pred new1(int,int,int).
:- pred diff1(int,int,int,int).
:- pred diff2(int,int,int,int).
:- pred new2(int,int,int).

false :- N1 =\= N2, new1(X,N1,N2).
new1(A,0,0).
new1(A,B,C) :- C=H+1, A=E, diff1(A,E,B,B1), new1(A,B1,H).
new1(A,B,C) :- A=\=E, diff2(A,E,B,B2), new1(A,B2,C).
diff1(A,A,1,0).
diff1(A,A,B1,B) :- B1=B+1, A=<C, new2(A,B,C).
diff1(A,A,B,C) :- D>E, A=E, diff1(A,E,B,C).
diff2(A,B,C1,D1) :- A>B, C1=C+1, D1=D+1, diff2(A,B,C,D).
diff2(A,B,0,0) :- A=\=B.
diff2(A,B,C,C) :- A=\=B, B=<D, new2(A,C,D).
diff2(A,B,C,D) :- A=\=B, A=\=E, E>B, diff2(A,B,C,D).
new2(A,B,C) :- A=C, B=1.
new2(A,B1,A) :- A=<C, B1=B+1, new2(A,B,C).
new2(A,B1,C) :- A>C, B1=B+1, new2(A,B,C).
new2(A,0,B) :- A=\=B.
new2(A,B,C) :- A=\=C, C=<E, new2(A,B,E).
new2(A,B,C) :- A=\=D, D>C, new2(A,B,C).
\end{verbatim}

\subsubsection{{\large{\rm Model of the transformed constrained Horn clauses.}}}

\begin{verbatim}
diff1(A,B,C,D) :- (((C - D) = 1), (C >= 1)).
diff2(A,B,C,D) :- ((C = D), (C >= 0)).
new1(A,B,C) :- ((B = C), (B >= 0)).
new2(A,B,C) :- (B >= 0).
\end{verbatim}

\subsection{{\rm Program} {\it SelectionSort} {\rm and property}
 {\it SS\_Orderedness.}}

\subsubsection{{\large{\rm Constrained Horn clauses obtained by translating 
{\it SelectionSort} and {\it SS\_Orderedness.}}}}

\begin{verbatim}
:- pred min(list(int),int).
:- pred delete(int,list(int),list(int)).
:- pred selectionSort(list(int),list(int)).
:- pred ordered(list(int),bool).

false :- selectionSort(L,S), ordered(S,false).
min([X],X).
min([X|T],M) :- X=<M1, M=X, min(T,M1).
min([X|T],M) :- X>M1, M=M1, min(T,M1).
delete(X,[Y|T],T) :- X=Y.
delete(X,[Y|T],[Y|D]) :- X=\=Y, delete(X,T,D).
selectionSort([],[]).
selectionSort(L,[M|T]) :- min(L,M), delete(M,L,L1), selectionSort(L1,T).
ordered([],true).
ordered([X],true).
ordered([X,Y|T],false) :- X>Y.
ordered([X,Y|T],B) :- X=<Y, ordered([Y|T],B).
\end{verbatim}

\subsubsection{{\large{\rm Transformed constrained Horn clauses.}}}

\begin{verbatim}
:- pred new1(bool).
:- pred new2(int,int).

false :- new1(false).
new1(true).
new1(false) :- new2(M,Y), M>Y.
new1(B) :- new1(B).
new2(M,Y) :- X>M, new2(M,Y), M>Y.
\end{verbatim}

\vspace*{-2mm}
\subsubsection{{\large{\rm Model of the transformed constrained Horn clauses.}}}

\begin{verbatim}
new1(A) :- (A = true).
new2(A,B) :- false.
\end{verbatim}

\vspace*{-2mm}
\subsection{{\rm Program} {\it SelectionSort} 
{\rm and property} {\it SS\_Length.}}

\subsubsection{{\large{\rm Constrained Horn clauses obtained by translating 
{\it SelectionSort} and {\it SS\_Length.}}}}

\begin{verbatim}
:- pred length(list(int),int).
:- pred min(list(int),int).
:- pred delete(int,list(int),list(int)).
:- pred selectionSort(list(int),list(int)).

false :- N1>=0, N2>=0, N1=\=N2,
    length(L,N1), selectionSort(L,S), length(S,N2).
length([],0).
length([X|Xs],N) :- N1>=0, N=N1+1, length(Xs,N1).
min([X],X).
min([X|T],M) :- X=<M1, M=X, min(T,M1).
min([X|T],M) :- X>M1, M=M1, min(T,M1).
delete(X,[Y|T],T) :- X=Y.
delete(X,[Y|T],[Y|D]) :- X=\=Y, delete(X,T,D).
selectionSort([],[]).
selectionSort(L,[M|T]) :- min(L,M), delete(M,L,L1), selectionSort(L1,T).
\end{verbatim}

\subsubsection{{\large{\rm Transformed constrained Horn clauses.}}}

\begin{verbatim}
:- pred new1(int,int).
:- pred diff(int,int,int).
:- pred new2(int,int).

false :- N1>=0, N2>=0, N1=\=N2, new1(N1,N2).
new1(0,0).
new1(N1,N2) :- N3>=0, N2=N3+1, new1(N4,N3), diff(X,N1,N4).
diff(H,1,0).
diff(X,N1,N2) :- N2>=0, N1=N2+1, X=<M1, new2(N2,M1).
diff(X,N1,N4) :- N2>=0, N1=N2+1, H>X, N5>=0, N4=N5+1, diff(X,N2,N5).
new2(1,H).
new2(N2,H)  :- N2=N5+1, H=<M2, new2(N5,M2).
new2(N2,M1) :- N2=N5+1, H>M1, new2(N5,M1).
\end{verbatim}

\subsubsection{{\large{\rm Model of the transformed constrained Horn clauses.}}}

\begin{verbatim}
diff(A,B,C) :- (((((B - C) = 1), (B >= 3)); ((B = 2), (C = 1))); 
    ((B = 1), (C = 0))).
new1(A,B) :- ((A = B), (A >= 0)).
new2(A,B) :- (A >= 1).
\end{verbatim}

\subsection{{\rm Program} {\it QuickSort} 
{\rm and property} {\it QS\_Perm.}}

\subsubsection{{\large{\rm Constrained Horn clauses obtained by translating 
{\it QuickSort} and {\it QS\_Perm.}}}}

\begin{verbatim}
:- pred append(list(int),list(int),list(int)).
:- pred partition(int,list(int),list(int),list(int)).
:- pred quickSort(list(int),list(int)).
:- pred count(int,list(int),int).

false :- N1 =\= N2, count(X,L,N1), quickSort(L,S), count(X,S,N2).
append([],Ys,Ys).
append([X|Xs],Ys,[X|Zs]) :- append(Xs,Ys,Zs).
partition(X,[],[],[]).
partition(X,[Y|Ys],[Y|L1],L2) :- Y=<X, partition(X,Ys,L1,L2).
partition(X,[Y|Ys],L1,[Y|L2]) :- X<Y, partition(X,Ys,L1,L2).
quickSort([],[]).
quickSort([X|Xs],S) :- partition(X,Xs,Ys,Zs),
   quickSort(Ys,S1), quickSort(Zs,S2), append(S1,[X|S2],S).
count(X,[],0).
count(X,[H|T],N) :- X=H, N=M+1, count(X,T,M).
count(X,[H|T],N) :- X=\=H, count(X,T,N).
\end{verbatim}

\subsubsection{{\large{\rm Transformed constrained Horn clauses.}}}

\begin{verbatim}
:- pred new1(int,int,int).
:- pred new2(int,int).
:- pred new3(int,int,int,int,int).
:- pred new4(int,int,int,int,int).
:- pred diff1(int,int,int,int,int).
:- pred diff2(int,int,int,int,int).
:- pred diff3(int,int,int,int,int).
:- pred diff4(int,int,int,int,int).

false :- M=\=N, new1(X,M,N).
new1(X,0,0).
new1(X,M,N) :- M=K+1, X=Y, diff1(X,Y,K,Ma,Mb), new1(X,Ma,Na),
    new1(X,Mb,Nb), diff2(X,Y,N,Na,Nb).
new1(X,M,N) :- X=\=Y, diff3(X,Y,M,Ma,Mb), new1(X,Ma,Na),
    new1(X,Mb,Nb), diff4(X,Y,N,Na,Nb).
diff1(X,Y,0,0,0)   :- X=Y.
diff1(X,Y,M,Ma,Mb) :- X=Y, Ma=Ma1+1, M=M1+1, diff1(X,Y,M1,Ma1,Mb).
diff2(X,Y,N,0,Nb)  :- X=Y, N=Nb+1, new2(X,Nb).
diff2(X,Y,N,Na,Nb) :- X=Y, Na=Na1+1, N=N1+1, diff2(X,Y,N1,Na1,Nb).
diff3(X,Y,M,Ma,Mb) :- X<Y, new3(X,Y,M,Ma,Mb).
diff3(X,Y,M,Ma,Mb) :- Y<X, new4(X,Y,M,Ma,Mb).
diff4(X,Y,N,0,N)   :- X=\=Y, new2(X,N).
diff4(X,Y,N,Na,Nb) :- X=\=Y, Na=Na1+1, N=N1+1, diff4(X,Y,N1,Na1,Nb).
new2(X,0).
new2(X,N) :- N=N1+1, new2(X,N1).
new3(X,Y,0,0,0)   :- X<Y.
new3(X,Y,M,Ma,Mb) :- X<Y, M=M1+1, Ma=Ma1+1, new3(X,Y,M1,Ma1,Mb).
new4(X,Y,0,0,0)   :- X>Y.
new4(X,Y,M,Ma,Mb) :- Y<X, M=M1+1, Mb=Mb1+1, new4(X,Y,M1,Ma,Mb1).
\end{verbatim}

\subsubsection{{\large{\rm Model of the transformed constrained Horn clauses.}}}

\begin{verbatim}
diff1(A,B,C,D,E) :- (((((((A = B), (C = 2)), (D = 2)), (E = 0));
  ((((A = B), (C = 1)), (D = 1)), (E = 0)));
  (((C = D), (E = 0)), ((((B + C) - A) >= 3), ((A + (C - B)) >= 3))));
  (((C = 0), (D = 0)), (E = 0))).
diff2(A,B,C,D,E) :- (((
  (((C - D) = 1), (((((B + C) - A) >= 2), ((A + (C - B)) >= 2)), (E >= 0)));
  ((((C - E) = 1), (D = 0)), ((((B + C) - A) >= 2), ((A + (C - B)) >= 2))));
  (((C = 1), (D = 0)), (E = 0)));
  (((((((B + C) - A) >= 2),
    (((B + D) - A) >= 1)), ((A + (C - B)) >= 2)), ((A + (D - B)) >= 1)), 
    (E >= 1))).
diff3(A,B,C,D,E) :- ((((((((C = D), (E = 0)), (C >= 3));
  (((C = E), (D = 0)), (C >= 2))); (((C = 2), (D = 2)),
  (E = 0))); (((C = 1), (D = 1)), (E = 0))); (((C = 1), (D = 0)), (E = 1)));
  (((C = 0), (D = 0)), (E = 0))).
diff4(A,B,C,D,E) :- ((((((((C = D), (E = 0)), (C >= 1));
  (((C = E), (D = 0)), (C >= 2)));
  (((C = 1), (D = 0)), (E = 1))); (((C = 0), (D = 0)), (E = 0)));
  ((E = 1), ((C >= 0), (D >= 1))));
  (((C >= 0), (D >= 1)), (E >= 2))).
new1(A,B,C) :- ((((B = C), (B >= 2)); ((B = 1), (C = 1))); ((B = 0), (C = 0))).
new2(A,B) :- (B >= 0).
new3(A,B,C,D,E) :- ((((((C = D), (E = 0)), (C >= 3)); 
  (((C = 2), (D = 2)), (E = 0)));
  (((C = 1), (D = 1)), (E = 0))); (((C = 0), (D = 0)), (E = 0))).
new4(A,B,C,D,E) :- (((((C = E), (D = 0)), (C >= 2)); 
  (((C = 1), (D = 0)), (E = 1))); (((C = 0), (D = 0)), (E = 0))).
\end{verbatim}

\subsection{{\rm Program} {\it QuickSort} {\rm and property} 
{\it QS\_Sum.}}

\subsubsection{\large{\rm Constrained Horn clauses obtained by translating 
{\it QuickSort} and {\it QS\_Sum.}}}

\begin{verbatim}
:- pred sumlist(list(int),int).
:- pred quickSort(list(int),list(int)).
:- pred partition(int,list(int),list(int),list(int)).
:- pred append(list(int),list(int),list(int)).

false :- N1 =\= N2, sumlist(L,N1), quickSort(L,S), sumlist(S,N2).
sumlist([],0).
sumlist([X|Xs],S) :- S=S1+X, sumlist(Xs,S1).
quickSort([],[]).
quickSort([D|T],S) :- partition(D,T,T1,T2), quickSort(T1,S1), 
    quickSort(T2,S2), append(S1,[D|S2],S).
partition(D,[],[],[]).
partition(D,[H|T],[H|L1],L2) :- H=<D,   partition(D,T,L1,L2).
partition(D,[H|T],L1,[H|L2]) :- D=<H-1, partition(D,T,L1,L2).
append([],Ys,Ys).
append([X|Xs],Ys,[X|Zs]) :- append(Xs,Ys,Zs).
\end{verbatim}

\subsubsection{{\large{\rm Transformed constrained Horn clauses.}}}

\begin{verbatim}
:- pred new1(int,int).
:- pred diff1(int,int,int,int).
:- pred diff2(int,int,int,int).

false :- new1(N1,N2), N1 =\= N2.
new1(0,0).
new1(N1,N2) :- N1=N3+H, diff1(N3,H,K1,K2), new1(K1,J1), new1(K2,J2), 
    diff2(H,N2,J1,J2).
diff1(0,H,0,0).
diff1(N3,H,K1,K2) :- N3=N4+X, H>=X, K1=K3+X, diff1(N4,H,K3,K2).
diff1(N3,H,K1,K2) :- N3=N4+X, H=<X-1, K2=K4+X, diff1(N4,H,K1,K4).
diff2(H,N2,0,N3)  :- N2=H+N3.
diff2(H,N2,J1,J2) :- N2=N3+Y, J1=J3+Y, diff2(H,N3,J3,J2).
\end{verbatim}

\subsubsection{{\large{\rm Model of the transformed constrained Horn clauses.}}}

\begin{verbatim}
diff1(A,B,C,D) :- ((A + ((-1 * C) - D)) = 0).
diff2(A,B,C,D) :- ((A + ((C + D) - B)) = 0).
new1(A,B) :- (A = B).
\end{verbatim}

\subsection{{\rm Program} {\it MergeSort} {\rm and property} 
{\it MS\_Sum.}}

\subsubsection{\large{\rm Constrained Horn clauses obtained by translating 
{\it MergeSort} and {\it MS\_Sum}.}}

\begin{verbatim}
:- pred sumlist(list(int),int).
:- pred merge(list(int),list(int),list(int)).
:- pred mergeSort(list(int),list(int)).
:- pred split(list(int),list(int),list(int)).
:- pred mydiv(int,int,int).
:- pred length(list(int),int).
:- pred takedrop(int,list(int),list(int),list(int)).
\end{verbatim}
\begin{verbatim}
false :- N1 =\= N2, sumlist(L,N1), mergeSort(L,S), sumlist(S,N2).
sumlist([],0).
sumlist([X|Xs],S) :- S=S1+X, sumlist(Xs,S1).
mergeSort([],[]).
mergeSort([H],[H]).
mergeSort([H|T],S) :- T=[Y|T1], split([H|T],L1,L2),
    mergeSort(L1,S1), mergeSort(L2,S2), merge(S1,S2,S).
\end{verbatim}
\begin{verbatim}
split(L,L1,L2) :- length(L,N), mydiv(N,2,N1), takedrop(N1,L,L1,L2).
mydiv(N,2,N1) :- N=N1+N2, N1-N2=<1.
length([],0).
length([H|T],N) :- N=M+1, length(T,M).
takedrop(N,[],[],[]).
takedrop(0,L,[],L).
takedrop(N,[H|T],[H|T1],L) :- N>=1, N1=N-1, takedrop(N1,T,T1,L).
merge([],L,L).
merge([X|Xs],[],[X|Xs]).
merge([X|Xs],[Y|Ys],[X|Zs]) :- X=<Y, merge(Xs,[Y|Ys],Zs).
merge([X|Xs],[Y|Ys],[Y|Zs]) :- X>=Y+1, merge([X|Xs],Ys,Zs).
\end{verbatim}

\subsubsection{{\large{\rm Transformed constrained Horn clauses.}}}

\begin{verbatim}
:- pred new1(int,int).
:- pred diff1(int,int,int,int,int).
:- pred diff2(int,int,int).
:- pred new2(int,int).
:- pred new3(int,int,int,int,int).
:- pred new4(int).
:- pred new5(int,int,int,int).
:- pred new6(int,int,int,int).

false :- new1(N1,N2), N1 =\= N2.
new1(0,0).
new1(H,H).
new1(N1,N2) :- N1=N3+X, diff1(Y,N3,X,K1,K2), new1(K1,J1),
    new1(K2,J2), diff2(N2,J1,J2).
diff1(A,B,C,0,D) :- B=E+A, H>=0, I=H+1, G>=0, H=G+1, I=0+J,
    0-J=<1, D=K+C, K=E+A, new2(E,G).
diff1(A,B,C,D,E) :- B=F+A, I>=0, J=I+1, H>=0, I=H+1, J=K+L,
    K-L=<1, K>=1, 0=K-1, D=0+C, E=F+A, new2(F,H).
diff1(A,B,C,D,E) :- B=F+A, J>=0, K=J+1, G>=0, J=G+1,
    K=L+M, L-M=<1, L>=1, N=L-1, N>=1, H=N-1, D=O+C, O=I+A.
diff2(A,0,B) :- new4(A), A=B, new3(F,G,H,I,E).
diff2(A,B,0) :- A=C+D, A=B, new4(C).
diff2(A,B,C) :- D=<E, A=F+D, B=G+D, C=H+E, new5(E,F,G,H).
diff2(A,B,C) :- D>=E+1, A=F+E, B=G+D, C=H+E, new6(D,F,G,H).
new2(0,0).
new2(C,D) :- C=E+A, F>=0, D=F+1, new2(E,F).
new3(0,0,A,0,0) :- A>=0.
new3(0,0,0,0,0) :- 0>=0.
new3(A,B,0,0,C) :- D>=0, B=D+1, A=F+G, C=F+G, new2(F,D).
new3(A,B,C,D,E) :- C>=0, F>=0, B=F+1, C>=1, H=C-1, H>=0,
    A=G+J, D=I+J, new3(G,F,H,I,E).
new4(0).
new4(A) :- A=B+C, new4(B).
new5(A,B,0,C) :- B=C+A, new4(C).
new5(A,B,C,D) :- E=<A, B=F+E, C=G+E, new5(A,F,G,D).
new5(A,B,C,D) :- E>=A+1, B=F+A, C=G+E, new6(E,F,G,D).
new6(A,B,C,0) :- B=C+A, new4(C).
new6(A,B,C,D) :- A=<E, B=F+A, D=G+E, new5(E,F,C,G).
new6(A,B,C,D) :- A>=E+1, B=F+E, D=G+E, new6(A,F,C,G).
\end{verbatim}

\subsubsection{{\large{\rm Model of the transformed constrained Horn clauses.}}}

\begin{verbatim}
diff1(A,B,C,D,E) :- ((B + (C + ((-1 * D) - E))) = 0).
diff2(A,B,C) :- ((A + ((-1 * B) - C)) = 0).
new1(A,B) :- (A = B).
new2(A,B) :- (B >= 0).
new3(A,B,C,D,E) :- (((A + ((-1 * D) - E)) = 0), ((B >= 0), (C >= 0))).
new4(A) :- true.
new5(A,B,C,D) :- ((A + ((C + D) - B)) = 0).
new6(A,B,C,D) :- ((A + ((C + D) - B)) = 0).
\end{verbatim}


\end{document}